     Set text width and height to reasonable size
   \newcommand{\be}[0]{\begin{equation}}
   \newcommand{\ee}[0]{\end{equation}}
   \newcommand{\ba}[0]{\begin{eqnarray}}
   \newcommand{\ea}[0]{\end{eqnarray}}    
\documentstyle[12pt]{article}
\begin{document}
\Large
\hfill\vbox{\hbox{DTP/97/46}
            \hbox{June 1997}}
\nopagebreak

\vspace{0.75cm}
\begin{center}
\LARGE
{\bf A Convergent Reformulation of QCD Perturbation Theory}
\vspace{0.6cm}
\Large

C.J. Maxwell

\vspace{0.4cm}
\large
\begin{em}
Centre for Particle Theory, University of Durham\\
South Road, Durham, DH1 3LE, England
\end{em}

\vspace{1.7cm}

\end{center}
\normalsize
\vspace{0.45cm}

\centerline{\bf Abstract}
\vspace{0.3cm}
%%%%%%%%%%%%ABSTRACT%%%%%%%%%%%%%%%%%%
We propose a generalization of Grunberg's method of
effective charges in which,  starting with the effective
charge for some dimensionless QCD observable
dependent on the single energy scale $Q$, ${\cal{R}}(Q)$,
 we introduce an infinite set of auxiliary
effective charges, each one describing the sub-asymptotic
$Q$-evolution of the immediately preceding effective charge.
The corresponding infinite set of coupled integrated effective
charge beta-function equations may be truncated. The 
resulting approximations for ${\cal{R}}(Q)$ are the convergents
of a continued function. They are manifestly RS-invariant
and converge to a limit equal to the Borel sum of the
standard asymptotic perturbation series in ${\alpha_s}({\mu^2})$,
with remaining ambiguities due to infra-red renormalons.
There are close connections with Pad{\'e} approximation.
\newpage

QCD perturbation theory in its conventional formulation
of a power series expansion in the renormalization group (RG)
improved coupling, ${\alpha_s}({\mu^2})$, suffers from a
number of defects. The series has a zero radius of convergence
with coefficients exhibiting factorial growth, and consequently
its sum can only be reconstructed by treating it as asymptotic
to some function of the coupling. Customarily Borel summation
is used for this reconstruction but of course this is not
the only possible choice, and the method to be used is not   
uniquely specified by the physical theory itself \cite{r1}.
Whilst ultra-violet (UV) renormalon singularities in the
Borel plane cause no problems in the definition of the Borel
sum, there are ambiguities resulting from infra-red (IR)
renormalons which, it has been argued , are closely connected
with, and can compensate, corresponding ambiguities in the
operator product expansion (OPE) \cite{r2}.

In addition to these problems arising from the large-order
behaviour of conventional perturbation theory, fixed-order
perturbation theory is afflicted with the problem of
renormalization scheme (RS) dependence. Thus fixed-order
truncations of the series depend on the renormalization 
convention used to define ${\alpha_s}({\mu^2})$. A number of
proposals for controlling or avoiding this RS-dependence 
problem have been advanced [3-6], but no consensus on the
issue has been reached.

In this paper we suggest a reformulation of QCD perturbation
theory which avoids many of these problems. In particular it
is convergent, notwithstanding remaining ambiguities due
to IR renormalons, and is manifestly RS-invariant. It can be
regarded as a generalization of Grunberg's effective charge
approach \cite{r4} and turns out to have a close relation
to Pad{\'e} approximation \cite{r7,r8}.

We begin by considering a generic dimensionless QCD observable
${\cal{R}}(Q)$, which depends on the single energy scale $Q$
(e.g. the ${e^+}{e^-}$ R-ratio with $Q$ the c.m. energy).
By raising to a power and scaling we can always arrange
that the formal perturbation series for ${\cal{R}}(Q)$ assumes
the form,

\be
{{\cal{R}}(Q)}=a+{r_1}{a^2}+{r_2}{a^3}+\ldots+{r_n}{a^{n+1}}
+\ldots\;,
\ee
where $a\equiv{\alpha_s}({\mu^2})/{\pi}$ 
is the RG-improved coupling
. We shall refer to observables defined in this way as
{\it effective charges} \cite{r4,r6}.

Such effective charges satisfy several important properties.
We shall consider SU($N$) QCD with $N_f$ flavours of massless
quark, and assume that the first beta-function coefficient
$b=(11N-2{N_f})/6$ is positive. The first familiar property
is Asymptotic Freedom (AF), which is equivalent to the
statement that for any effective charge ${\cal{R}}(Q)$,

\be
\lim_{Q\rightarrow{\infty}}{\cal{R}}(Q)={0}\;.
\ee
The second property, which we shall refer to as ``Asymptotic
Scaling'' (AS) , is perhaps less familiar, and will be central
to the reformulation of perturbation theory we are proposing.
The first step is to define a universal QCD scaling
function,

\be
{\cal{F}}(x)\equiv{e}^{-1/bx}{(1+1/cx)}^{c/b}\;,
\ee
where $c$ is the second (universal) beta-function coefficient,

\be
c=\left[-\frac{7}{8}\frac{{C_A}^2}{b}-\frac{11}{8}\frac{{C_A}{C_F}}
{b}+\frac{5}{4}{C_A}+\frac{3}{4}{C_F}\right]\;,
\ee
with ${C_A}=N$, ${C_F}=({N^2}-1)/2N$ , SU($N$) Casimirs.
Then AS corresponds to the statement that for any effective
charge ${\cal{R}}(Q)$,

\be
\lim_{Q\rightarrow{\infty}}Q{\cal{F}}({\cal{R}}(Q))={\Lambda}_
{\cal{R}}\;,
\ee
where ${\Lambda}_{\cal{R}}$ is a scaling constant with
dimensions of energy which depends on the observable.
Given sufficiently large values of $Q$ this property can
evidently serve as a test of QCD, but since ${\Lambda}_
{\cal{R}}$ is not universal it cannot usefully be applied    
at fixed values of $Q$ . However if the next-to-leading order
(NLO) perturbative coefficient $r\equiv{r}_{1}^{\overline{MS}}
({\mu}=Q)$ has been calculated (in the $\overline{MS}$ scheme
with scale ${{\mu}^2}={Q^2}$ for instance), then
${\Lambda}_{\cal{R}}$ can be converted into a 
{\it universal} scaling constant ${\Lambda}_{\overline{MS}}$
via the {\it exact} Celmaster-Gonsalves \cite{r9} relation,
 \be
{\Lambda}_{\cal{R}}{e}^{-r/b}={\Lambda}_{\overline{MS}}\;.
\ee
Use of different subtraction procedures results in different
constants which, however, are still {\it universal}. We 
stress that $r$ is independent of $Q$. Thus we arrive at the
property of ``Universal Asymptotic Scaling'' (UAS),

\be
\lim_{Q\rightarrow{\infty}}Q{\cal{F}}({\cal{R}}(Q))
{e}^{-r/b}={\Lambda}_{\overline{MS}}\;.
\ee

This property can now be used to test QCD at fixed $Q$ by
looking at the scatter in $Q{\cal{F}}({\cal{R}}(Q)){e}^{-r/b}$
for various observables \cite{r6}. At finite $Q$ UAS will be
violated and one will have

\be
Q{\cal{F}}({\cal{R}}(Q)){e}^{-r/b}={\Lambda}_{\overline{MS}}
(1+{\eta}_{\cal{R}}(Q))\;,
\ee
where ${\eta}_{\cal{R}}(Q)$ represents the non-scaling
sub-asymptotic effects which vanish as $Q\rightarrow{\infty}$
. The crucial observation is that ${\eta}_{\cal{R}}(Q)$ has
the formal perturbation series

\be
{\eta}_{\cal{R}}(Q)=\frac{{\rho_2}}{b}(a+O({a^2}))\;,
\ee
where ${\rho_2}$ is the next-NLO (NNLO) effective charge (EC)
beta-function coefficient \cite{r4,r6},
given by the RS-invariant combination ,

\be
{\rho_2}={c_2}+{r_2}-{r_1}{c}-{{r_1}^2}\;,
\ee
with $c_2$ the three-loop beta-function coefficient. Thus
${\eta}_{\cal{R}}$ is proportional to {\it another} effective
charge, which will itself satisfy the UAS property, and will
have its own non-scaling contribution proportional to
{\it yet another} effective charge, etc.  This self-similar
construction based on the exact property of UAS provides the
reformulation of QCD perturbation theory which we are
proposing.

The construction can usefully be regarded as a generalization
of Grunberg's effective charge approach \cite{r4,r6}, and
it will help to illuminate the discussion so far if we 
briefly review that formalism. We refer the reader to
reference [6] for full details.

The key ingredient in the EC approach is the EC beta-function
${\rho}({\cal{R}}(Q))$ which determines the $Q$-evolution
of the observable ${\cal{R}}(Q)$,
    
\ba
\frac{d{\cal{R}}(Q)}{d\ln Q}&\equiv&-b{\rho}({\cal{R}}(Q))
\nonumber\\
&=&-b({{\cal{R}}^2}+{c}{{\cal{R}}^3}+{\rho_2}{{\cal{R}}^4}
+\ldots+{\rho_n}{{\cal{R}}^{n+2}}+\ldots)\;,
\ea
where the ${\rho_n}$ are $Q$-independent RS-invariant
combinations of the $r_i$, $c_i$ 
($i\leq{n}$) (see eq.(10) for ${\rho_2}$). ${\rho}({\cal{R}}(Q))$ 
can be regarded as an observable to be reconstructed from the
measured running of ${\cal{R}}(Q)$ \cite{r6}. Integrating up
eq.(11) and imposing Asymptotic Freedom as a boundary
condition yields,

\ba
{F}({\cal{R}}(Q))&=&b\ln{\frac{Q}{{\Lambda}_{\cal{R}}}}-
\int^{{\cal{R}}(Q)}_{0}{dx}\left[-\frac{1}{\rho(x)}+
\frac{1}{{x^2}(1+cx)}\right]\nonumber\\
&\equiv&b\ln{\frac{Q}{{\Lambda}_{\cal{R}}}}-
{\Delta}{\rho_0}(Q)\;,
\ea
where

\be
F(x)\equiv-b\ln{{\cal{F}}(x)}=\frac{1}{x}+c\ln\left(\frac{cx}
{1+cx}\right)\;,
\ee
and with ${\Lambda}_{\cal{R}}$ a constant of integration. Assuming
AF the AS property in eq.(5) then follows directly on
rearranging eq.(12) and taking the $Q\rightarrow{\infty}$ limit,
in which ${\Delta}{\rho_0}(Q)\rightarrow{0}$ . Further
, as demonstrated in ref.[6], one can then identify
$b\ln(Q/{{\Lambda}_{\cal{R}}})$ with the NLO RS-invariant
${\rho_0}(Q)$ \cite{r3,r4},

\be
b\ln{\frac{Q}{{\Lambda}_{\cal{R}}}}={\rho_0}(Q)
\equiv{b}\ln{\frac{Q}{{\Lambda}_{\overline{MS}}}}-r\;,
\ee
from which the Celmaster-Gonsalves relation in eq.(6) follows.
The sub-asymptotic ${\eta}_{\cal{R}}(Q)$ contribution in
eq.(8) is equivalent to $(\exp({{\Delta}{\rho_0}/b})-1)$,
with ${\Delta}{\rho_0}(Q)$ as defined in eq.(12).

In this notation the violation of UAS is controlled by
${\Delta}{\rho_0}(Q)$ . Expanding the integrand in eq.(12)
as a power series in $x$ and integrating term-by-term we
obtain

\be
{\Delta}{\rho_0}(Q)={\rho_2}{\cal{R}}+({\rho_3}-2{\rho_2}{c})
\frac{{\cal{R}}^2}{2}+({\rho_4}-2{c}{\rho_3}+2{c^2}{\rho_2}
+2{c^2}-{{\rho_2}^2})\frac{{\cal{R}}^3}{3}+\ldots\;.
\ee
We can then write ${\Delta}{\rho_0}(Q)\equiv{\rho_2}
{{\cal{R}}^{(1)}}(Q)$, where ${\cal{R}}^{(1)}(Q)$ is
another effective charge whose perturbative expansion
follows on substituting ${\cal{R}}=a+{r_1}{a^2}+\ldots$
in eq.(15),

\be
{\cal{R}}^{(1)}(Q)=a+({r_1}+\frac{\rho_3}{2{\rho_2}}-c){a^2}
+\ldots\;,
\ee
from which the coefficients ${r}_{1}^{(1)}$,${r}_{2}^{(1)}$,
$\ldots$, can be obtained. ${\cal{R}}^{(1)}(Q)$ will satisfy
eq.(12) with ${\rho_0}(Q)$, ${\Delta}{\rho_0}(Q)$ replaced
by ${\rho}_{0}^{(1)}(Q)$ and ${\Delta}{\rho}_{0}^{(1)}(Q)$
where

\be
{\rho}_{0}^{(1)}(Q)=b\ln \frac{Q}{{\Lambda}_{\overline{MS}}}-
{r}_{1}^{(1){\overline{MS}}}={b}\ln\frac{Q}{{\Lambda}_
{\overline{MS}}}-r-\frac{\rho_3}{2{\rho_2}}+c\;.
\ee
Writing ${\Delta}{\rho}_{0}^{(1)}(Q)\equiv{\rho}_{2}^{(1)}
{\cal{R}}^{(2)}(Q)$ with ${\cal{R}}^{(2)}(Q)$ yet another
effective charge the self-similar construction continues.

Thus perturbation theory is reformulated as an infinite
set of effective charges ${\cal{R}}(Q)$,${\cal{R}}^{(1)}(Q)$,
$\ldots$,${\cal{R}}^{(n)}(Q)$,$\ldots$, satisfying the
set of coupled equations,

\ba
F({\cal{R}}(Q))&=&{\rho_0}(Q)-{\rho_2}{\cal{R}}^{(1)}(Q)
\nonumber\\
F({\cal{R}}^{(1)}(Q))&=&{\rho}_{0}^{(1)}(Q)-
{\rho}_{2}^{(1)}{\cal{R}}^{(2)}(Q)\nonumber\\
F({\cal{R}}^{(2)}(Q))&=&{\rho}_{0}^{(2)}(Q)-{\rho}_{2}^{(2)}
{\cal{R}}^{(3)}(Q)\nonumber\\
\vdots & & \vdots
\nonumber\\
F({\cal{R}}^{(n)}(Q))&=&{\rho}_{0}^{(n)}(Q)-{\rho}_{2}^{(n)}
{\cal{R}}^{(n+1)}(Q)\nonumber\\
\vdots & &\vdots
\ea
${\rho}_{0}^{(n)}$ involves $b\ln(Q/{{\Lambda}_{\overline{MS}}})$,
$r$, and the ${\rho}_{k}$ EC RS invariants for ${k}\leq{2n+1}$
 (we define ${\rho}_{1}\equiv{c}$). Thus it is determined
given a complete ${\rm N}^{2n+1}$LO perturbative calculation for
${\cal{R}}(Q)$ . ${\rho}_{2}^{(n)}$ involves the ${\rho}_k$
for ${k}\leq{2n+2}$ and so is determined given a
${\rm N}^{2n+2}$LO calculation for ${\cal{R}}(Q)$.

The effective charge ${\cal{R}}^{(1)}(Q)$ describes the
violation of UAS of ${\cal{R}}(Q)$, ${\cal{R}}^{(2)}(Q)$
describes the scaling violation of ${\cal{R}}^{(1)}(Q)$ etc.
If we assume that ${\cal{R}}^{(n)}(Q)$ scales and set
${\cal{R}}^{(n+1)}(Q)=0$, then the truncated set of
equations in eqs.(18) can be solved for ${\cal{R}}(Q)$,
given $\ln({Q}/{{\Lambda}_{\overline{MS}}})$, the NLO
$\overline{MS}$ coefficient $r$, and the ${\rho}_{0}^{(i)}$,
${\rho}_{2}^{(i)}$, that is given a complete   
${\rm N}^{2n+1}$LO perturbative calculation for ${\cal{R}}(Q)$.

Equivalently we can iteratively eliminate ${\cal{R}}^{(1)}$,
${\cal{R}}^{(2)},\ldots$ from eqs.(18). If we define $G(x)$
to be the inverse function of $F(x)$ in eq.(13), that is
$G(F(x))=x$, then we obtain the continued function
representation for ${\cal{R}}(Q)$,

\be
{\cal{R}}(Q)=G({\rho_0}-{\rho_2}G({{\rho}_{0}^{(1)}}-
{{\rho}_{2}^{(1)}}G({{\rho}_{0}^{(2)}}-{{\rho}_{2}^{(2)}}G
(\ldots)))\ldots)\;.
\ee

Equation (19) is our main result. The first NLO approximation
$G({\rho}_{0})$ is precisely the conventional NLO perturbative
approximation in the EC scheme. Subsequent approximations
differ from conventional fixed-order perturbation theory.
An interesting feature of the approach is that each extra
level in the construction of eqs.(18) adds {\it two} further
orders in perturbation theory. One can interpolate by
using $G({\rho_0}-{\rho_2}G({\rho_0}))$ as the NNLO
approximation, and then the ${\rm N}^3$LO approximation
is $G({\rho_0}-{\rho_2}G({{\rho}_{0}^{(1)}}))$,
$G({\rho_0}-{\rho_2}G({{\rho}_{0}^{(1)}}-{{\rho}_{2}^{(1)}}G
({{\rho}_{0}^{(1)}})))$ interpolates the ${\rm N}^4$LO, etc.

Our claim is that if only ultra-violet (UV) renormalon
singularities are present then this sequence of approximations
converges, and the limit is precisely the Borel sum of
the divergent asymptotic conventional perturbation series
in ${\alpha_s}({\mu^2})$. If infra-red (IR) renormalons are
present in addition then strict convergence is prevented
\cite{r8}. Since by construction eqs.(18) are a set of
{\it exact} equations this indicates that these equations
are insufficient to determine ${\cal{R}}(Q)$ and
correspondingly extra information must be added, presumably in
the form of further OPE terms.

We shall now motivate this convergence claim by considering
the simplified case of QCD with $c=0$. In this limit
$F(x)=G(x)=1/x$ and the continued function of eq.(19) becomes
a continued fraction.

\be
{\cal{R}}(Q)=\frac{1|}{|{\rho}_{0}}-\frac{{\rho_2}|}
{|{\rho}_{0}^{(1)}}-\frac{{\rho}_{2}^{(1)}|}{|{\rho}
_{0}^{(2)}}-\ldots\;.
\ee
As we shall now show the successive convergents of this
continued fraction are the diagonal $[n/n]$ Pad{\'e} approximants
of the original perturbation series for ${\cal{R}}(Q)$ in
eq.(1), in the RS where ${c_2}={c_3}=\ldots={c_k}=\ldots=0$,
the so-called `t Hooft scheme \cite{r10}. This is in accord
with the recent observation by Gardi that in the ``large-
${\beta}_{0}$'' limit diagonal Pad{\'e} approximants are
RS-invariant \cite{r7}.   

The diagonal $[n/n]$ Pad{\'e} approximants are defined by

\ba
{\cal{R}}[n/n]&\equiv&\frac{a+{A_2}{a^2}+\ldots+{A_n}{a^n}}
{1+{B_1}{a}+\ldots+{B_n}{a^n}}\nonumber\\
&=& a+{r_1}{a^2}+{r_2}{a^3}+\ldots+{r}_{2n-1}{a}^{2n}+
O({a}^{2n+1})\;,
\ea
where the $A_i$ and $B_i$ coefficients are
uniquely fixed by demanding that the ${r_i}, {i}\leq{2n-1}$
are reproduced on expanding the quotient.

Corresponding to the original perturbation series in
eq.(1) one can define the standard Stieltjes \cite{r11}
continued fraction for ${\cal{R}}(Q)$,

\be
{\cal{R}}(Q)=\frac{{K_1}{a}|}{|1}+\frac{{K_2}{a}|}{|1}
+\frac{{K_3}{a}|}{|1}+\ldots\;,
\ee
where ${K_1}=1,{K_2}=-{r_1},{K_3}={r_1}-{r_2}/{r_1},\ldots$.
The successive convergents of eq.(22) are respectively the
diagonal $[n/n]$ and off-diagonal $[n/n+1]$ Pad{\'e}
approximants to ${\cal{R}}(Q)$. We can also define the
so-called ``associated'' or Jacobi form of the continued
fraction \cite{r11},

\be
{\cal{R}}(Q)=\frac{{K_1}{a}|}{|(1+{K_2}{a})}-
\frac{{K_2}{K_3}{a^2}|}{|1+({K_3}+{K_4}){a}}-
\frac{{K_4}{K_5}{a^2}|}{|1+({K_5}+{K_6}){a}}-\ldots\;.
\ee
The convergents of eq.(23) are the diagonal $[n/n]$ Pad{\'e}
approximants of ${\cal{R}}$. If we scale the partial 
quotients in eq.(23) by $a$ we obtain

\be
{\cal{R}}(Q)=\frac{{K_1}|}{|({1}/{a})+{K_2}}-
\frac{{K_2}{K_3}|}{|({1}/{a})+{K_3}+{K_4}}-
\frac{{K_4}{K_5}|}{|({1}/{a})+{K_5}+{K_6}}-\ldots\;.
\ee
In the `t Hooft scheme (${c_2}={c_3}=\ldots=0$), and with
$c=0$ , ${{\rho}_{0}^{(n)}}=(1/a)-{r}_{1}^{(n)}$, and so
on comparing eq.(24) with eq.(20) we can identify
${{\rho}_{0}^{(n)}}=(1/a)+{K}_{2n+1}+{K}_{2n+2}$ and
${{\rho}_{2}^{(n)}}={{K}_{2n+2}}{{K}_{2n+3}}$, and we
see that indeed the successive (manifestly RS-invariant)
convergents of eq.(20) are precisely the successive convergents
of the Jacobi form of the continued fraction- the diagonal
${\cal{R}}[n/n]$ Pad{\'e} approximants.

The convergence of the diagonal Pad{\'e} approximants is assured if 
the perturbation series for ${\cal{R}}(Q)$ is asymptotic to
a ``Stieltjes
function'' \cite{r11,r12}, and if this is the case this
function is identical to the Borel sum of the series, if it
exists \cite{r12}.

A Stieltjes function is of the form \cite{r12}

\be
{\cal{R}}(a)=a\int_{0}^{\infty}{dt} \frac{{W}(t)}{(1+ta)}\;,
\ee
where ${W}(t)$ is nonnegative throughout the range of
integration. For such a function the continued fraction
coefficients $K_n$ are all nonnegative \cite{r12}.

A relevant example, motivated by UV renormalon singularities
\cite{r1}, is to consider a simple pole at $z=-1$ in
the Borel plane,

\be
{\cal{R}}(a)=\int_{0}^{\infty}{dz}\,{e}^{-z/a}\frac{1}{(1+z)}
=a\int_{0}^{\infty}{dt}\frac{{e}^{-t}}{(1+ta)}\;,
\ee
so that the non-negative Stieltjes weight function in
eq.(25) is ${W}(t)={e}^{-t}$. This corresponds to
perturbative coefficients ${r_n}={(-1)}^{n}n!$ which results
in the (nonnegative) continued fraction coefficients
${K_1}=1,{K}_{2n}={K}_{2n+1}=n,n>{1}$, from which we can
obtain ${\rho}_{0}^{(n)}=(1/a)+2n+1$ and ${\rho}_{2}^{(n)}
={(n+1)}^{2}$ for the coefficients in eqs.(18). The
continued fraction in eq.(20) then converges to the
well-defined Borel sum in eq.(26). Branch point
UV renormalon singularities can also be reduced to Stieltjes
form by a change of variables. A subtlety is that if one
changes the RS , and hence the definition of $a$, the series
may no longer be Stieltjes, equivalently the $K_n$ are
RS-dependent and will not all be nonnegative in a general RS.
To prove convergence, however, it is only necessary that the 
series is Stieltjes in one {\it particular} RS, since the
convergents are RS-invariant.

By analyzing the large-$n$ behaviour of eqs.(18) using
${\rho}_{0}^{(n)}\approx{2n}$ and ${\rho}_{2}^{(n)}\approx
{n^2}$ one can infer that for the above example
${\cal{R}}^{(n)}(Q)\approx{1/n}$, so as ${n}\rightarrow
{\infty}$, ${\cal{R}}^{(n)}(Q)\rightarrow{0}$, and successive
violations of UAS become ever smaller, underwriting the
convergence of the truncations of eqs.(18).

To conclude, we regard the important feature of this approach
to be its physical motivation based on iterating an exact 
property of massless QCD, which we have termed Universal
Asymptotic Scaling, and which is defined in eq.(7). Thus
if the infinite set of coupled equations in eqs.(18) has a
solution this determines ${\cal{R}}(Q)$ . It is incidental
that this approach can reproduce the result of applying
Borel summation to the original divergent asymptotic series.
For instance if ${r_n}={(-1)}^{n}(2n)!$ the Borel method
is inapplicable, nonetheless the set of coupled equations
eqs.(18) in this case generates a continued fraction
(assuming $c=0$) which converges to a unique Stieltjes
function \cite{r12}.

We plan to give a more complete discussion of the convergence
of the approach, and the manner in which ambiguities induced by
the presence of IR renormalons manifest themselves in this
language, in a future work. The links with Pad{\'e} methods
are also interesting and should be explored further.

\section*{Acknowledgements}

We thank Jan Fischer for a number of enjoyable discussions.
The CERN TH Division
is thanked for its hospitality during the period when this
work was completed.

\end{document}